# Comparison of the Crystal Structure of the Heavy-Fermion Materials CeCoIn$_5$, CeRhIn$_5$ and CeIrIn$_5$

E. G. Moshopoulou[1*], J. L. Sarrao[2], P. G. Pagliuso[2], N. O. Moreno[2], J. D. Thompson[2], Z. Fisk[2,3], R. M. Ibberson[4]

[1]National Center for Scientific Research "Demokritos", Institute of Materials Sciences, 15310 Agia Paraskevi, Greece

[2]Condensed Matter and Thermal Physics, MS K764, Los Alamos National Laboratory, Los Alamos, NM 87545, USA

[3]National High Magnetic Field Laboratory, Florida State University, Tallahassee, FL32306, USA

[4]ISIS Facility, Rutherford Appleton Laboratory, Chilton, Didcot, Oxon OX11, 0QX, United Kingdom

(February 6, 2002)

## Abstract

The crystal structure of the recently discovered heavy-fermion (HF) superconductor CeCoIn$_5$ ($T_c$ = 2.3 K) has been determined by high-resolution neutron powder diffraction. It is tetragonal (space group P4/*mmm*), with lattice parameters $a$ = 4.61292(9) Å and $c$ = 7.5513(2) Å at ambient conditions. Whereas CeCoIn$_5$ is isostructural with the HF aniferromagnet CeRhIn$_5$ and the HF superconductor CeIrIn$_5$, its cell constants and its only variable positional parameter, $z$In2, differ significantly from the corresponding ones of CeRhIn$_5$ and CeIrIn$_5$. As a result, the distortions of the cuboctahedron [CeIn$_3$], which is the key structural unit in all three materials, are different in CeCoIn$_5$ from the ones in CeRhIn$_5$ and CeIrIn$_5$. The compounds CeCoIn$_5$ and

CeIrIn$_5$, which contain the most distorted (in one or another way) [CeIn$_3$] cuboctahedra exhibit superconductivity at ambient pressure below 2.3 K and 0.4 K respectively. On the other hand, CeRhIn$_5$, in which [CeIn$_3$] cuboctahedra are the less distorted, and the parent compound of the materials CeTIn$_5$ i.e. the cubic HF CeIn$_3$ are antiferromagnets at ambient pressure with $T_N$ = 3.8 K and 10 K correspondingly.

Pacs 61.12.Ld, 71.27.+a, 72.15.Qm, 74.70.Tx

*evagelia@ims.demokritos.gr, tel.+30-1-6503320; fax +30-1-6519430

1. INTRODUCTION

A recent advance in the physics and chemistry of heavy-fermion (HF) systems has been the discovery of the quasi-2D materials $CeTIn_5$, where T = Co, Rh, Ir. Collectively, these new systems enhance our knowledge of superconductivity, magnetism and HF behavior, and provide a suitable experimental environment to search for possible structure-property relationships in HF materials.

The systems $CeTIn_5$ exhibit fascinating and unexpected physical properties. $CeRhIn_5$ is a HF antiferromagnet with electronic coefficient of specific heat $\gamma \equiv C/T \geq 420$ mJ/(molCe $K^2$) and Néel temperature $T_N = 3.8$ K. Application of hydrostatic pressure of about 16 kbar induces a first-order like transition from antiferromagnetic state to an unconventional superconducting state with $T_c = 2.1$ K (ref. 1). This transition is not expected within widely accepted today theoretical predictions and is markedly different from any previously reported for a HF material. $CeCoIn_5$ is a HF superconductor at ambient pressure with Sommerfeld coefficient $\gamma \approx 290$ mJ/(molCe $K^2$) at 2.4 K and critical temperature $T_c = 2.3$ K which is the highest known $T_c$ for a HF system (ref. 2). $CeIrIn_5$ is also a HF superconductor at ambient pressure with $\gamma \approx 750$ mJ/(molCe $K^2$) and a bulk superconducting transition temperature $T_c = 0.4$ K. An intriguing feature of $CeIrIn_5$ is that its resistivity drops to zero at $T_0 = 1.2$ K without any obvious thermodynamic signature (ref. 3). Specific heat and thermal conductivity (ref. 4, 5) as well as NMR experiments (ref. 6, 7) showed that the pairing symmetry in the superconducting state of $CeCoIn_5$ and $CeIrIn_5$ is unconventional.

Single crystal neutron, x-ray and electron diffraction experiments on $CeRhIn_5$ and $CeIrIn_5$ (ref. 8) and conventional powder x-ray diffraction on CeCoIn$_5$ (ref. 2, 9, 10) demonstrated that the compounds CeTIn$_5$ adopt the tetragonal HoCoGa$_5$-type structure (s.g. P4/*mmm*). The compounds are built by monolayers of face-sharing distorted cuboctahedra [CeIn$_3$] and monolayers of edge-sharing rectangular parallelepipeds [TIn$_2$], stacked alternatively in the [001] direction. The key structural unit of the series is the distorted cuboctahedron [CeIn$_3$]. Such structural arrangement implies that CeCoIn$_5$, CeRhIn$_5$ and CeIrIn$_5$ are quasi-2D variants of CeIn$_3$. CeIn$_3$ is a cubic HF antiferromagnet in which superconductivity is induced by pressure of ≈ 25 kbar and below $T_c$ = 200 mK (ref. 11). Whether the reduced dimensionality is responsible for the significant increase of $T_c$ in the materials CeTIn$_5$ is an outstanding question, which, if answered, will have an important impact on the present understanding of the physics of superconductivity not only in the HF materials but also in the broader class of correlated electron systems.

Our current structural investigations of the materials CeTIn$_5$ at both ambient and non-ambient conditions aim to determine accurately their detailed crystallographic parameters and reveal the structural aspects associated with the rich array of phases and phenomena described above. In this conference, we present the crystal structure of CeCoIn$_5$ at ambient conditions and compare it with those of CeRhIn$_5$, CeIrIn$_5$ and CeIn$_3$ reported earlier.

2. DATA COLLECTION AND ANALYSIS

Time-of-flight neutron powder diffraction data were collected at the High Resolution Powder Diffractometer (HRPD) at ISIS facility of Rutherford Appleton Laboratory. The data were collected in the backscattering detector bank with Δd/d resolution of ~ 4×10$^{-4}$. This unprecedented resolution is effectively constant over the whole d-spacing range 0.75 to 2.35 Å where the data were collected. It allows a highly accurate determination of the crystal structure as well as detection of potential very subtle changes in symmetry.

Single crystals of $CeCoIn_5$ were ground into a fine powder and packed loosely in a flat plate holder; such sample geometry is the optimal one in order to reduce absorption effects. Local ISIS programs were used for the instrument control, data acquisition and corrections of the data from instrumental effects. The data analysis and structure refinements were carried out by GSAS (ref. 12). The initial structural model was the structure of $CeRhIn_5$ (ref. 8), i.e. s.g. P4/*mmm*, *a* = *b* = 4.656 Å, *c* = 7.542 Å and atomic coordinates: Ce at (0, 0, 0), Co at (0, 0, 1/2), In1 at (1/2, 1/2, 0) and In2 at (0, 1/2, $z$ = 0.306). The parameters varied in the refinement were the scale factor, ten background coefficients, four profile-coefficients, the lattice constants, the positional parameter $zIn2$, nine thermal factors, three diffractometer constants and the preferred orientation, extinction and absorption coefficients. The detailed procedure for the data analysis and refinement of the structure as well as the values of all refined parameters will be reported in detail elsewhere. The Rietveld refinement fit is given in Fig. 1. The cell constants, the positional parameter and the main interatomic distances of $CeCoIn_5$ are listed in Table 1 and compared with the corresponding ones of $CeRhIn_5$ and $CeIrIn_5$ obtained from reference 8, as well as with the respective ones of $CeIn_3$ obtained from reference 13. It is worth noting here that the crystallographic parameters of $CeIrIn_5$ were also deduced from high resolution powder diffraction data taken at HRPD and they are in excellent agreement with the parameters reported in reference 8, which were resulted from single crystal diffraction data.

## 3. DISCUSSION

As can be seen in Table 1, in $CeCoIn_5$ the interatomic distances Ce-In1 are smaller than the Ce-In2 ones. This means that the $[CeIn_3]$ cuboctahedra in $CeCoIn_5$ are distorted (compared with the cubic $CeIn_3$) differently than the ones in $CeRhIn_5$ and $CeIrIn_5$. In the former compound the chemical pressure induced on each $[CeIn_3]$ by its two adjacent $[CoIn_2]$ rectangular parallelepipeds, is lower than the chemical pressure due to its four adjacent $[CeIn_3]$ cuboctahedra of the same layer. As a result, the $[CeIn_3]$ cuboctahedra of $CeCoIn_5$ are slightly elongated along the c axis, while they are dilated in the (a, b) plane in $CeRhIn_5$ and $CeIrIn_5$.

Comparison of the structural parameters of $CeCoIn_5$ given in Table 1, with the recently reported ones by Settai et al (ref. 10) reveal an important difference: the parameters reported in this reference, obtained from Rietveld analysis of conventional powder x-ray diffraction data, result in undistorted $[CeIn_3]$ cuboctahedra. Better (but not complete) agreement exists between the results of the present study and the ones of Kalychak et al. (9) obtained again from conventional powder x-ray diffraction data but on arc-melted samples.

Finally, we point out that the $[CeIn_3]$ cuboctahedra are more distorted (in one or another way) in the two superconducting compounds $CeCoIn_5$ and $CeIrIn_5$ than in the antiferromagnet $CeRhIn_5$. Possibly, the stronger distortions in $CeCoIn_5$ and $CeIrIn_5$ generate special features in their underlying electronic structure and magnetic fluctuation spectra, which are at the origin of superconductivity in these systems. Band structure calculations as well as high-resolution inelastic neutron scattering experiments are necessary to address this issue. Undoubtedly, the structural trend just outlined suggests that some structural parameters of the CeTIn$_5$ and CeIn$_3$ can be correlated with their physical properties; however a more general and relevant structure-property relationship can be obtained only through comparison of the structural responses of the compounds CeTIn$_5$ and CeIn$_3$ around their corresponding critical points.

*Acknowledgements*: E. G. M. acknowledges the support of the European Community, Access to Research Infrastructure Action of the Improving Human Potential Program. Work at Los Alamos National Laboratory was performed under the auspices of the US Department of Energy. Z. F. acknowledges partial support from U.S. National Science Foundation Grant No. DMR-9971348.

CAPTIONS

Table 1: Select Structural and Refinement parameters for CeIn$_3$, CeCoIn$_5$, CeRhIn$_5$, CeIrIn$_5$

Fig. 1: Rietveld Refinement fit of CeCoIn$_5$

Table 1

|  | CeIn$_3$ | CeCoIn$_5$ | CeRhIn$_5$ | CeIrIn$_5$ |
|---|---|---|---|---|
| $a$(Å) | 4.689(2) | 4.61292(9) | 4.656 (2) | 4.674(1) |
| $c$(Å) |  | 7.5513(2) | 7.542(1) | 7.501(5) |
| $z$In2 |  | 0.3094(3) | 0.3059(2) | 0.3052(2) |
| Ce-In1 × 4 (Å) | 3.3156(6) | 3.26183(6) | 3.292(2) | 3.3050(7) |
| Ce-In2 × 8 (Å) |  | 3.283(1) | 3.2775(7) | 3.272(1) |
| T-In × 8 (Å) |  | 2.7187(9) | 2.7500(9) | 2.7560(7) |
| $R$ |  | 0.0572 | 0.056 | 0.051 |
| $R_w$ |  | 0.0669 | 0.1152 | 0.082 |
| $\chi^2$ |  | 1.133 | 1.674 | 1.552 |

Fig. 1

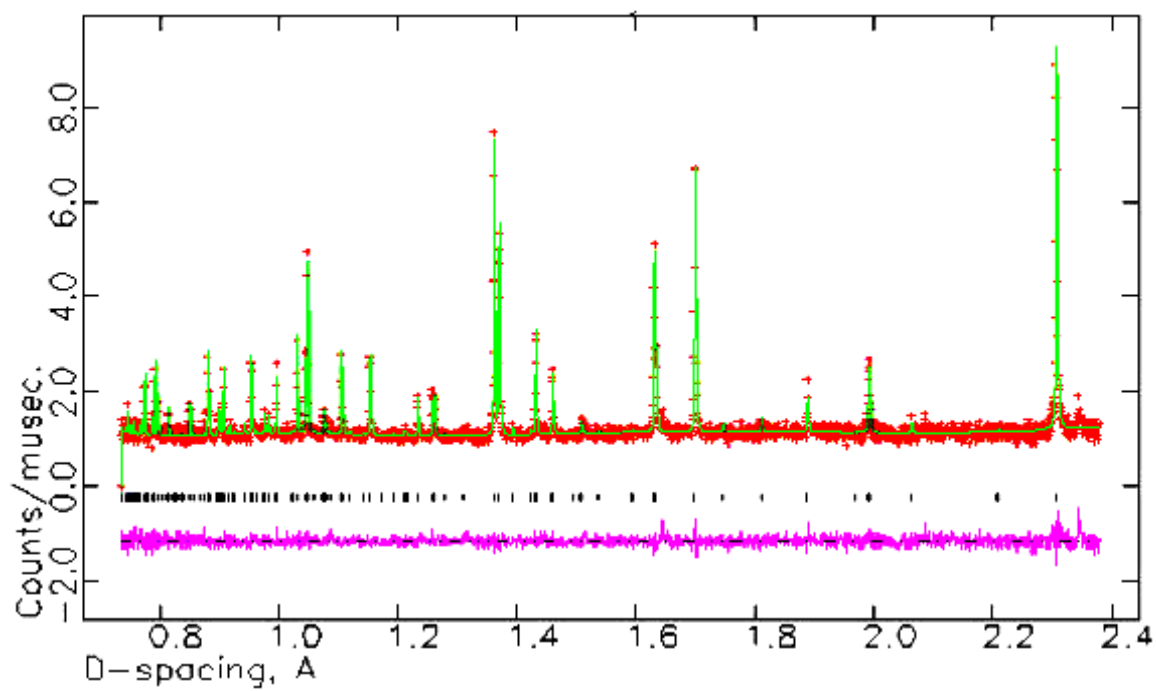